\documentclass[prb,twocolumn,showpacs,showkeys]{revtex4}

\usepackage{amssymb}

\usepackage{graphicx}% Include figure files
\usepackage{dcolumn}% Align table columns on decimal point
\usepackage{bm}
\usepackage{color}
\begin{document}

\title{Signatures of asymmetric and inelastic tunneling on the spin torque bias dependence}
\date{\today}
\author{A. Manchon$^{1,2}$,S. Zhang$^{2}$,K.-J. Lee$^{3}$}
\affiliation{$^1$ Division of Physical Science and Engineering, KAUST, Thuwal 23955, Saudi Arabia;\\$^2$Department of Physics, University of Arizona, Tucson,
AZ 85721, USA;\\$^3$Department of Materials Science and Engineering, Korea University, Seoul 136-713, Korea.}
\begin{abstract}
The influence of structural asymmetries (barrier height and exchange splitting), as well as inelastic scattering (magnons and phonons) on the bias dependence of the spin transfer torque in a magnetic tunnel junction is studied theoretically using the free electron model. We show that they modify the "conventional" bias dependence of the spin transfer torque, together with the bias dependence of the conductance. In particular, both structural asymmetries and bulk (inelastic) scattering add {\em antisymmetric} terms to the perpendicular torque ($\propto V$ and $\propto j_e|V|$ ), while the interfacial inelastic scattering conserves the junction symmetry and only produces {\em symmetric} terms ($\propto |V|^n$, $n\in\mathbb{N}$). The analysis of spin torque and conductance measurements displays a signature revealing the origin (asymmetry or inelastic scattering) of the discrepancy.
\end{abstract}
\maketitle
%\clearpage

\section{Introduction}

The recent observation of current-driven magnetization control \cite{Slonc96} in Magnetic Tunnel Junctions \cite{huai} (MTJs) offers promising opportunities for magnetic recording and memories applications \cite{fullerton}. The observed critical switching current has now reached 10$^6$A/cm$^2$, which makes MTJs competitive candidates for Magnetic Random Access Memories \cite{ieee}.\par

However, due to the specific transport properties in MTJs, the characteristics of the spin transfer torque in these devices display significant differences with the current-driven torque usually observed in metallic spin-valves \cite{sunralph}. Uncovering the precise form of the bias dependence of the spin torque is essential to understand and control the dynamical properties of the magnetization. In MTJs, it has been demonstrated both theoretically \cite{slonc07,theo,kioussis,manchon,xiao,wil,heiliger} and experimentally \cite{sankey,deac,petit,li,sun,oh} that the torque possesses two components, usually referred to as the in-plane (or Slonczewski) torque, $T_{||}$ and the perpendicular (or out-of-plane) torque, $T_\bot$. The first one is purely non-equilibrium and competes with the damping, whereas the second one arises from spin reorientation at the interfaces, possesses both equilibrium (Interlayer Exchange Coupling \cite{slonc89}) and non-equilibrium components and acts like an effective magnetic field on the magnetization. The presence of this perpendicular torque results in original dynamical properties of the magnetization \cite{sankey,deac,petit,li,sun,oh}.\par

Up until now, most of the experimental efforts have been focused on the bias-dependence of the perpendicular torque $T_\bot$. Although this component is vanishingly small in metallic spin-valves, it can not be neglected in MTJs, due to the momentum filtering imposed by the tunnel barrier \cite{manchon}. Most of the theories, using tight-binding \cite{theo,kioussis}, free-electron \cite{manchon,xiao,wil} or {\em ab-initio} \cite{heiliger} calculations, have addressed the bias dependence of the spin torque within a symmetric and purely elastic tunneling junction (referred to as SE tunneling). It has been shown that for SE tunneling at low bias voltage, the form of the spin torque is:
\begin{eqnarray}\label{eq:SEip}
{\bf T}_{||}&=&(a_1V+a_2V^2){\bf M}\times({\bf M}\times {\bf P}),\\\label{eq:SEop}
{\bf T}_{\bot}&=&(b_0+b_2V^2){\bf M}\times {\bf P},
\end{eqnarray}
where ${\bf P}$ and ${\bf M}$ are the magnetization directions of the pinned and free layers, respectively. These bias dependencies have been well observed in spin-diode-type experiments \cite{sankey} performed on MgO-based MTJs. The linear bias-dependence of the in-plane torque that has been measured ($a_2\approx0$) is consistent with Ref. \onlinecite{heiliger} which suggests that MgO-based MTJs behave like half-metallic junctions. \par

In contrast, a number of experiments using dynamical and switching properties of the MTJs \cite{deac,petit,li,sun,oh}, as well as recent theoretical investigations \cite{kioussis, xiao, wil} have recently questioned the validity of the "conventional" bias-dependencies represented by Eqs.(\ref{eq:SEip})-(\ref{eq:SEop}). In particular, Xiao et al. \cite{xiao} and Wilczynski et al. \cite{wil} employing the free electron model numerically showed that structural asymmetries could alter the convention bias dependence of the perpendicular torque, whereas Tang et al. \cite{kioussis} predicted a non-monotonic bias dependence of $T_\bot$, demonstrating the importance of band filling. From the experimental side, Li et al. \cite{li} measured a field-like effect of the form $\propto j_e|V|$ and interpreted their data by considering the electron-magnon scattering in the bulk of the ferromagnets. In contrast, Sun et al. \cite{sun} suggested the possibility of non macrospin processes or heating artefacts that would induce a bias-dependent effective field. Very recenlty, Oh et al. \cite{oh} demonstrated the possibility to tune the bias dependence of the perpendicular torque by engineering the structural asymmetries of a MgO-based MTJ, consistently with theoretical simulations \cite{xiao,wil,kioussis}. One of the authors also proposed that an incomplete absorption of the transverse spin density within the free layer could lead to an asymmetric perpendicular torque \cite{manchon}. Finally, we recently studied the influence of interfacial electron-magnon scattering on the bias dependence of the torque \cite{magnon} and found that an additional symmetric term of the form $\propto|V|$ arises.\par

As seen from this brief overview, the bias dependence of the spin torque is far from universal, and a number of mechanisms has been shown to modify this dependence. However, the role of asymmetries has been investigated numerically within the tight binding model \cite{kioussis} and the
free electron model \cite{xiao,wil} and little is known concerning the role of inelastic scattering \cite{li,magnon}. In this paper, we derive analytic
solutions for $T_{||}$ and $T_{\bot}$ in the case of structural asymmetries and (bulk and interfacial) inelastic scattering by magnons and phonons, leading to a discrepancy between the actual torques and the "conventional" ones [Eqs. (1)-(2)]. In particular, both structural asymmetries and bulk (inelastic) scattering add {\em antisymmetric} terms to the perpendicular torque ($\propto V$ and $\propto j_e|V|$ ), while the interfacial inelastic scattering conserves the junction symmetry and only produces {\em symmetric} terms ($\propto |V|$). Moreover, we suggest that a connection exists between the tunneling conductance and the out-of-plane torque which constitutes a signature of the origin of the discrepancy.\par

This paper is organized as follows: in section II, we briefly discuss the form of the tunneling spin torque with and without spin diffusion. Section III addresses the bias-dependence of the spin torque and conductance in the presence of structural asymmetries. The influence of bulk and interfacial inelastic scattering is described in section IV and the conclusion is given in section V.

\section{Spin current vs spin density}

In most of the theoretical studies on the spin transfer torque in MTJs, the torques are associated with the transverse spin current density at the interfaces between the electrodes and the tunnel barrier \cite{slonc07,theo,kioussis,manchon,xiao,wil,heiliger}. This definition is only valid in the case of semi-infinite electrodes where the spin diffusion is neglected. A more correct approach is to relate the spin torque to the spin density rather than to the spin current (see Ref. \onlinecite{chapter} for a detailed discussion). The spin torque is then the torque exerted by the transverse spin density on the local magnetization and has the form:
\begin{eqnarray}
{\bf T}=\int_\Omega \frac{J}{\hbar}{\bf m}\times{\bf M}d\Omega,
\end{eqnarray}
where $J$ is the $s-d$ exchange coupling, ${\bf m}$ is the itinerant spin density, ${\bf M}$ is the local magnetization and $\Omega$ is the volume of the magnetic layer. The spin density can be computed from the well-known spin continuity equation \cite{chapter}:
\begin{eqnarray}\label{eq:eq6}
\frac{\partial {\bf m}}{\partial t}=-\nabla\cdot{\cal J}_s-\frac{J}{\hbar}{\bf m}\times{\bf M}-\frac{{\bf m}}{\tau_{sf}},
\end{eqnarray}
where ${\cal J}_s$ is the spin current and $\tau_{sf}$ is the spin relaxation time. In the case of a magnetic tunnel junction, where the resistance is dominated by the barrier, the spatial variation of the spin density is usually neglected and the torque is directly related to the spin current\cite{slonc07,theo,kioussis,manchon,xiao,wil,heiliger}: 
${\bf T}=-\int_\Omega \nabla\cdot{\cal J}_sd\Omega$. Therefore, in the case of a semi-infinite magnetic layer, the torque reduces to the {\em interfacial transverse spin current}.\par

However, in realistic junctions, the free layer is usually thin ($t\approx$2-3nm) and the torque arising at the interface between the barrier and the ferromagnetic electrode must be balanced by the torque arising at the second interface: ${\bf T}={\cal J}_s(x=0)-{\cal J}_s(x=t)$, which may introduce some deviations from the "conventional" bias dependence of the torque \cite{manchon}. On the other hand, in MgO-based junctions, the junction behaves like a half-metallic MTJ \cite{heiliger} and the spin density (or transverse spin current) is strongly absorbed near the barrier interface \cite{manchon,heiliger} (a few monolayers). Therefore, the usual identification ${\bf T}={\cal J}_s(x=0)$ is essentially valid in MgO-MTJs if one neglects the spin diffusion in the electrodes.\par

Nevertheless, we will show that it is possible to account for the spin relaxation ($1/\tau_{sf}\neq0$ in Eq. (\ref{eq:eq6})) in the bulk of the electrodes (impurities- or magnons-induced spin-flip scattering) as long as this relaxation does not significantly modify the interfacial densities of states, and thereby the tunneling process itself. In this case, we find that the resulting spin torque is a mixing between the two transverse components of the spin current. This issue will be addressed in detailed in section IV.

\section{Structural asymmetries\label{s:as}}

In Ref. \onlinecite{oh}, the authors demonstrated the possibility to add a linear component to the bias dependence of the perpendicular torque by intentionally introducing structural asymmetries in the junction. Depending on the asymmetry, it is possible to change the sign of the linear component, therefore artificially tuning the form of the spin torque. This finding is consistent with numerical studies \cite{kioussis,xiao,wil}. Although a connection is suggested between the bias dependence of the conductance and the one of the perpendicular torque \cite{oh, kioussis,xiao,wil}, this connection remains unclear and analytical formulae are needed.\par

In this section, we study the influence of two types of structural asymmetries. First, we consider the presence of different exchange splittings in the ferromagnetic electrodes. The exchange splittings of Fe, Co and Ni have been measured experimentally near the $\Gamma$ point \cite{Eastman}, as shown in Table 1. As a consequence, varying the composition of the electrodes, one can obtain different exchange splittings up to $J_R-J_L\approx0.5$ eV.\par

\begin{table}
	\centering
		\begin{tabular}{c|ccc}
			&Fe&Co&Ni\\\hline\hline
			$J$ \cite{Eastman} (eV)&1.5&1.1&0.6\\
			W \cite{Michaelson} (eV)&4.67-4.81&5&5.04-5.35
		\end{tabular}\caption{Exchange splitting and work functions for the three standard ferromagnetic transition metals.}\label{table:T1}
\end{table}

Another type of structural asymmetry is the presence of a different barrier height at the left and right interfaces of the junction. Since the work functions of Co, Fe and Ni are different \cite{Michaelson} (see Table \ref{table:T1}), the asymmetry can be created by using different electrode compositions, but also by modifying the composition of the barrier itself \cite{asym}.\par
\begin{figure}
  \includegraphics[width=8cm]{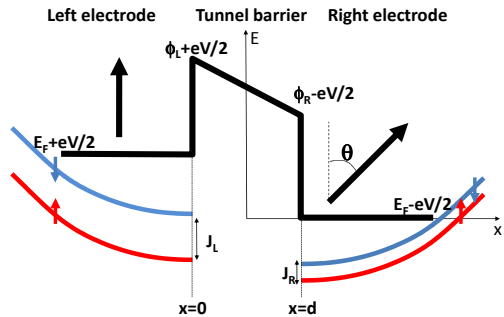}\\
   \caption{Potential profile of an asymmetric Magnetic Tunnel Junction. The right and left parabolae represent the dispersion of tunneling electrons.}\label{fig:fig1}
\end{figure}

We consider the junction presented in Fig. 1, where two ferromagnetic electrodes are separated by an insulator. The magnetizations form an angle $\theta$ between them. The barrier of average height $\phi=(\phi_R+\phi_L)/2$ possesses an asymmetry $\Delta\phi=\phi_R-\phi_L$, whereas the electrodes have an average exchange splitting $J=(J_R+J_L)/2$ with an asymmetry $\Delta J=J_R-J_L$. To determine the influence of these asymmetries on the spin torque and conductance, we use the same approach as Brinkman et al. \cite{brinkman}. Using the free electron model within the Keldysh formalism developed in Ref. \onlinecite{manchon}, the wave functions are determined for the complete structure (see Ref. \onlinecite{manchon}). The analytic forms of the torque and current are obtain up to the first order in $\exp[-2d\kappa_0]$, where $d$ is the barrier thickness and $\kappa_0=\sqrt{2m\phi/\hbar^2}$ is the barrier wave vector for perpendicularly incident Fermi electrons (see Appendix A). The effective mass of the electrons within the barrier is taken equal to 1. Therefore, the general form of the torques and current is:
\begin{equation}
T_{||},\;T_{\bot},\;J_e=\int\int dEd{\bf k}_{||}e^{-2d\kappa(E,{\bf k}_{||})}F(E,{\bf k}_{||}),
\end{equation}
where $F(E,{\bf k}_{||})$ is a function given explicitly in Appendix B, $E$ is the electron energy and ${\bf k}_{||}$ is the wave vector component in the plane of the layers. The factor $e^{-2d\kappa(E,{\bf k}_{||})}$ arises from the WKB approximation and represents the tunneling transmission. Following the spirit of Brinkman et al. \cite{brinkman}, we assume that the barrier is thick and high enough so that the energy dependence is essentially contained in the exponential factor $e^{-2d\kappa(E,{\bf k}_{||})}$. Therefore, $F(E,{\bf k}_{||})\approx F(E_F\pm eV/2,0)$, and we obtain:
\begin{widetext}
\begin{eqnarray}\label{eq:analtip}
T_{||}&=&T_{||0}\left[a_1\frac{eV}{\phi}+a_2\left(\frac{eV}{\phi}\right)^2\right]\sin\theta,\\\label{eq:analtop}
T_{\bot}&=&T_{\bot0}\left[1+b_1\frac{eV}{\phi}+b_2\left(\frac{eV}{\phi}\right)^2\right]\sin\theta,\\\label{eq:analg}
G_p(V)&=&G^p_0\left[1+g^p_1\frac{eV}{\phi}+g^p_2\left(\frac{eV}{\phi}\right)^2\right],\\
G_{ap}(V)&=&G^{ap}_0\left[1+g^{ap}_1\frac{eV}{\phi}+g^{ap}_2\left(\frac{eV}{\phi}\right)^2\right],
\end{eqnarray}
\end{widetext}
at the second order in bias voltage $V$. The torques $T_{||}$, $T_\bot$ are exerted on the {\em right} layer and $G_{p,ap}(V)$ is the conductance defined as $G_{p,ap}(V)=\partial J^{p,ap}_e/\partial V$ where $J^{p,ap}_e$ is the charge current in the parallel and antiparallel configurations, respectively. The coefficients $a_1...g_2^{ap}$ are given explicitly in Appendix C. Notice that up to the first order in the barrier, the angular dependence of the in-plane and perpendicular torques is a simple $\sin\theta$. The introduction of asymmetries does not modify the angular dependence of the torque, as long as the barrier is either high enough or thick enough ($\beta=d\kappa_0>>1$). \par

To illustrate the influence of the structural asymmetries on the torques and conductance, the analytical expressions given in Eqs. (\ref{eq:analtip})-(\ref{eq:analg}) have been plotted in Fig. 2, together with the full numerical simulation of the model developped in Ref. \onlinecite{manchon}. The torques and conductance are represented in their reduced form: the in-plane torque is normalized to the in-plane torquance ($\partial T_{||}/\partial V$) in the absence of asymmetries, whereas the perpendicular torque and conductance are normalized to their value at zero bias. Several points are worth noting.

First, since the in-plane torque is already asymmetric against the bias voltage in SE tunneling ($a_1$ and $a_2$ do not vanish in the absence of structural asymmetries), the "conventional" bias dependence given in Eq. (\ref{eq:SEip}) is conserved in the presence of asymmetries and only the actual magnitude of $a_1$ and $a_2$ is modified, as illustrated in Fig. 2(a,b). Note that the small discrepancy between the numerical model (solid lines) and the analytical expressions (squares) can be attributed to the presence of a cubic term $\propto V^3$ in the torque $T_{||}$. The change in the slope of the torque can be simply understood by considering the polarization (defined as Slonczewski's polarization - see Appendix B) of the electrons responsible for the in-plane torque $T_{||}$.\par

\begin{figure}
  \includegraphics[width=8cm]{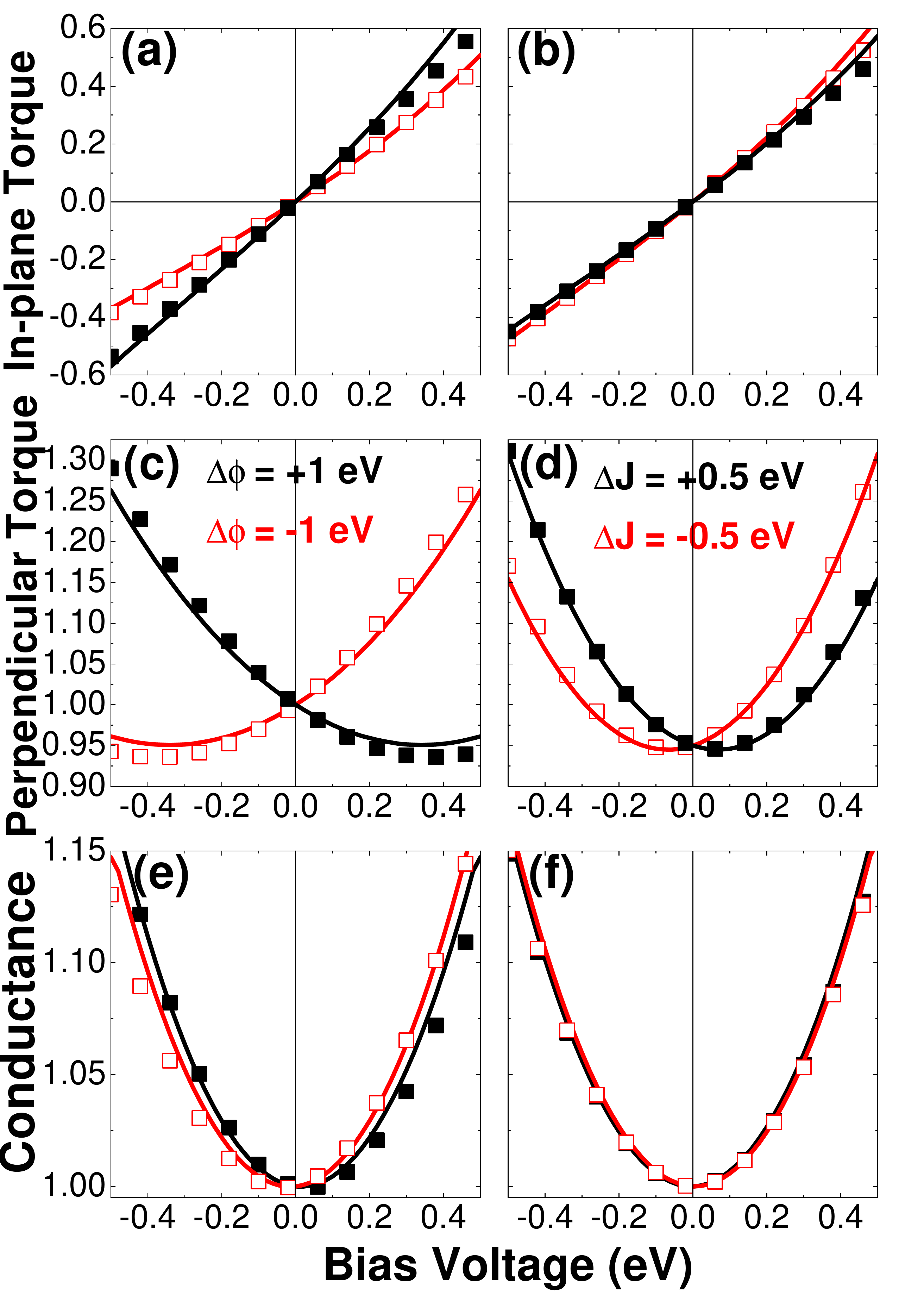}\\
  \caption{Bias dependence of the in-plane torque (a,b), perpendicular torque (c,d) and parallel conductance (e,f) in the case of barrier height (a,c,e) and exchange splitting (b,d,f) asymmetries. The solid lines correspond to numerical calculations based on the model presented in Ref. \onlinecite{manchon} and the squares are calculations using Eqs. (\ref{eq:analtip})-(\ref{eq:analg}). The parameters are $E_F=10$eV, $J=1$eV, $\phi=5$eV, $d=1$nm.)}\label{fig:fig2}
\end{figure}

Secondly, the terms $b_1$ and $g^{p,ap}_1$ are proportional to $\Delta J$ and $\Delta \phi$ so that in the absence of structural asymmetry, the perpendicular torque and the conductance are {\em quadratic} in bias voltage. However, when structural asymmetries are present, the perpendicular torque and the conductance both acquire an
additional {\em linear} component ($b_1$ and $g^{p,ap}_1$). This is consistent with numerical simulations reported earlier \cite{kioussis, xiao,wil} and the analytical expressions satisfactorily reproduce the numerical results, as shown in Fig. \ref{fig:fig2}(c-f).\par

An interesting feature here is the sign of the deviations.
For $\Delta\phi>0$ and $\Delta J=0$, the
junction is more conductive for negative bias ($\phi_R>\phi_L$), therefore a shift is
observed in the conductance and torque towards {\em positive} voltages ($b_1,g^p_1<0$ - see Fig. 2(a) and Fig. 2(c)). However,
in the case $\Delta J>0$ and $\Delta\phi=0$, the tunneling from
left to right is slightly more efficient (since $J_R>J_L$) and the
conductance displays a shift towards {\em negative} voltages ($g^p_1>0$ - see Fig. 2(d)). In contrast, the
electrons from the right electrodes are more polarized than the ones
from the left electrode and the torque displays a shift towards
{\em positive} voltages ($b_1<0$ - see Fig. 2(b)). This difference in the signature of the structural
asymmetry allows for the identification of the source
of the linear term in the out-of-plane torque, as demonstrated by the study of Oh et al. \cite{oh}.\par

These results are consistent with previous numerical studies \cite{kioussis, xiao, wil} at low bias. However, the comparison with the tight-binding model studied in Ref. \onlinecite{kioussis} presents some differences. The free electron model yields an open parabolic band dispersion whereas the tight-binding model produces a closed band dispersion. Therefore, the free electron model is only correct for low bias dependence and provides results for {\em low band filling}. As a consequence, the free electron model is surprisingly well adapted to the case of Fe. It also implies that the bias voltage must be smaller than the half-band width of the conduction electrons. As a consequence, neither band filling-induced sign reversal of IEC nor the oscillatory bias dependence \cite{kioussis} of the perpendicular torque can be obtained within the free electron model. The above results are limited to reasonably small bias and low band filling systems.

\section{Inelastic scattering\label{s:ine}}

In this section, we consider that the (bulk or interfacial) scattering by phonons, magnons or impurities in the left and right electrodes are symmetric (i.e., the interactions have the same amplitude in the left and right electrodes). This way, the electron scattering conserves the symmetry of the junction (it may not be true when the electrodes compositions are different). Although the symmetry of the system is conserved, the spin torque does not have the same expression in the case of bulk or interfacial scattering. As mentioned in section II, in the case of interfacial scattering the spin torque is directly related to the interfacial spin current, ${\bf T}={\cal J}_s(x=0)$, whereas in the case of bulk scattering, the spin relaxation in the electrodes cannot be neglected anymore and ${\bf T}={\cal J}_s(x=0)-\int_\Omega d\Omega{\bf m}/\tau_{sf}$. Therefore, although the symmetry of the MTJ is conserved in both cases, the bias dependence of the spin torque will experience a different modification depending on whether the scattering occurs at the interfaces or in the bulk of the electrodes.

\subsection{Interfacial scattering}

We consider two types of interfacial inelastic scattering processes: electron-magnon and electron-phonon. The influence of electron-magnon scattering on TMR \cite{zhang97,bratkovsky} and spin transfer torque \cite{magnon,swstt} has been studied within the Transfer Hamiltonian formalism. The current density is expressed in the form of a $2\times2$ spinor matrix:
\begin{eqnarray}\label{eq:5}
\hat{J}=2\pi\frac{e}{\hbar}&&\sum_{\bf{k,p}}[\hat{\rho}_{\bf{k}}\hat{T}_{\bf{kp}}\hat{\rho}_{\bf{p}}(\hat{T}_{\bf{kp}})^+f_L(1-f_R)\nonumber\\&&-\hat{\rho}_{\bf{p}}(\hat{T}_{\bf{kp}})^+\hat{\rho}_{\bf{k}}\hat{T}_{\bf{kp}}f_R(1-f_L)],
\end{eqnarray}
where $f_{L(R)}$ and ${\hat \rho}_{{\bf k}({\bf p})}$ are the Fermi distribution function and electronic density of states at the left (right) interfaces, and ${\hat T}_{\bf{kp}}$ (${\hat T}_{\bf{pk}}$) is the spin-dependent transfer matrix accounting for both elastic and inelastic tunneling. In the spinor formalism, the charge current and spin current are expressed $j_e=Tr[{\hat J}]$ and $T_{||(\bot)}={\cal J}^{x(y)}_s=Tr[{\hat \sigma}_{x(y)}{\hat J}]$, where ${\hat \sigma}_{x(y)}$ are the Pauli spin matrices.\par

In the case of electron scattering by interfacial phonons, although no spin-flip takes place, the increase of the conductance is expected to modify the TMR\cite{bratkovsky} and, correspondingly, the spin torque. In the presence of electron-magnon and electron-phonon interactions, the transfer matrix can be written:
\begin{eqnarray}\label{eq:sd}
\hat{T}_{{\bf kp}}^{e-m}&=&\hat{T}^d_{{\bf kp}}\left(\hat{I}+\sqrt{\frac{Q^m}{N}}(\bm{\sigma}.{\bf S}_{tr}^R+\bm{\sigma}.{\bf S}_{tr}^L)\right),\\\label{eq:ph}
\hat{T}_{{\bf kp}}^{e-ph}&=&\hat{T}^d_{{\bf kp}}\left(1+\sqrt{\frac{Q^{ph}_{\bf q}}{N}}(b_{\bf q}+b_{\bf q}^+)\right)\hat{I},
\end{eqnarray}
where $\hat{T}^d_{{\bf kp}}$ is the direct tunneling matrix, $Q^m$ ($Q^{ph}_{\bf q}$) is the phenomenological electron-magnon (electron-phonon)
efficiency, $N$ is the number of atoms per cell, $\bm{\sigma}$ is
the vector of Pauli spin matrices and ${\bf S}_{tr}^{L(R)}$ are the
transverse part of the magnetizations of the left and right
electrodes. Details about the derivation of Eq. (\ref{eq:sd}) can be found in Ref. \onlinecite{magnon}. The interaction efficiency ($Q^m$ and $Q^{ph}_{\bf q}$) can be related to quantum mechanical quantities, $Q^m\approx J^2/E^2_F$ and $Q^{ph}_{\bf q}\approx|M_{\bf q}^2|/E_F^2$, where $J$ is the exchange splitting, $E_F$ is the Fermi energy and $M_{\bf q}$ is the electron-phonon interaction \cite{phonons} that depends on the type of coupling (acoustic, optical or polar coupling).\par
\subsubsection{Electron-Phonon Scattering}
In the case of electron-phonon scattering, the transfer matrix Eq. (\ref{eq:ph}) is diagonal in spin space and obviously, the presence of phonons does not induce spin-flip. However, the direct tunneling matrix is renormalized by $(1+\sqrt{\frac{Q^{ph}_{\bf q}}{N}}(b_{\bf q}+b_{\bf q}^+))$ and becomes bias dependent \cite{bratkovsky}. We then expect that the modification of the conductance due to phonons alters the bias dependence of the spin torque.
Performing the matrix products displayed in Eq. (\ref{eq:5}) and using the definition of the spinor current stated above, we find:
\begin{widetext}
\begin{eqnarray}
j_e(E,{\bf q})&=&\frac{G_0}{e}\left(1+Q_{\bf q}^{ph}\langle b^+_{\bf q}b_{\bf q}\rangle+Q_{\bf q}^{ph}\langle b_{\bf q}b^+_{\bf q}\rangle\right)(1+\cos\theta P^LP^R)(f_L(1-f_R)-f_R(1-f_L)),\\
T_{||}(E,{\bf q})&=&\frac{G_0}{e}P_L\left(1+Q_{\bf q}^{ph}\langle b^+_{\bf q}b_{\bf q}\rangle+Q_{\bf q}^{ph}\langle b_{\bf q}b^+_{\bf q}\rangle\right)(f_L(1-f_R)-f_R(1-f_L))\sin\theta,\\
T_\bot(E,{\bf q})&=&\frac{G_0}{e}\left(1+Q_{\bf q}^{ph}\langle b^+_{\bf q}b_{\bf q}\rangle+Q_{\bf q}^{ph}\langle b_{\bf q}b^+_{\bf q}\rangle\right)(P^R\varphi_Lf_L(1-f_R)+P^L\varphi_Rf_R(1-f_L))\sin\theta,
\end{eqnarray}
\end{widetext}
where $P_{L,R}$ is the polarization at the left (right) interface and $\varphi_{R,L}$ is a coefficient that accounts for the spin rotation during tunneling \cite{magnon}. The integration rules are described in Ref. \onlinecite{magnon}. Assuming that the electron spin-dependent densities of state do not vary much over the range $eV$, and considering acoustic phonons ($\omega \propto q$, $Q_{\bf q}\propto q$) with a density of states of the form $\rho_{ph}(\omega)\propto \omega^{\nu}$, we obtain, at T=0 K and low bias voltage:
\begin{eqnarray}
&&G(V)=G_0(1+P^LP^R\cos\theta)(1+\zeta_{ph}|V|^{\nu+2}),\label{eq:16}\\
&&T_{||}=G_0P^L\sin\theta(1+\zeta_{ph}|V|^{\nu+2})V,\label{eq:17}\\
&&T_\bot-T_{\bot0}=G_0P^R\phi_L\sin\theta\zeta_{ph}|V|^{\nu+3},\label{eq:18}
\end{eqnarray}
$\zeta_{ph}$ being a coefficient that depends on the electron-phonon coupling, Fermi energy, Debye temperature $\Theta_D$ etc... The bias-dependence of the conductance ($\propto |V|^{\nu+2}$) is consistent with the one suggested by Bratkovsky when $\nu=2$. At larger bias, $|V|^{\nu+2}$ is replaced by $k_B\Theta_D$ and the bias dependence of the torques becomes linear. Note that the symmetry of the out-of-plane torque against the bias is conserved, whereas the in-plane torque acquires an antisymmetric component. \par

At higher temperature and bias, more complex behavior are found, but the bias dependence of $G(V)$ and $T_\bot$ is always an even function of $V$ ($|V|^n$, $n\in\mathbb{N}$). As an illustration, we provide below the expressions for large bias at finite temperature ($eV>k_BT>k_B\Theta_D$):
\begin{eqnarray}
&&G(V)= G_0(1+P^LP^R\cos\theta)\left(1+\xi_{ph}\frac{T}{\Theta_D}\left(\frac{|eV|}{k_B\Theta_D}\right)^{\nu+1}\right),\label{eq:phTG}\\
&&T_{||}= G_0\sin\theta P^L\left(1+\xi_{ph}\frac{T}{\Theta_D}\left(\frac{|eV|}{k_B\Theta_D}\right)^{\nu+1}\right)V,\label{eq:phTTip}\\
&&T_\bot-T_{\bot0}= G_0P^R\phi_L\sin\theta\xi_{ph}\frac{T}{\Theta_D}\left(\frac{|eV|}{k_B\Theta_D}\right)^{\nu+2}.\label{eq:phTTop}
\end{eqnarray}
Again $\xi_{ph}$ depends on the electron-phonon coupling, Fermi energy, Debye temperature etc. At finite temperatures, the conductance is enhanced, due to phonon-assisted tunneling and therefore, both in-plane and out-of-plane torques are enhanced. The temperature dependence is expected to be linear. Notice that although the magnitude of the torque increases with the temperature, its efficiency (ratio between spin torque and current density) is not modified, since the electron-phonon interaction does not affect the spin itself, but rather the tunneling rate.
\subsubsection{Electron-Magnon Scattering}
In the case of electron-magnon interaction, the transfer matrix [Eq. (\ref{eq:sd})] possesses non-diagonal elements that are responsible for spin-flip scattering. We then expect a much more complex influence on the torque. Assuming a magnon density of states of the form $\rho_m(\omega)=\omega^\nu$, symmetric electrodes ($P^L=P^R=P$, $\phi_L=\phi_R=\phi$) and T=0 K, we find:
\begin{eqnarray}
&&G(V)\propto (1-P^2\cos\theta)|V|^{\nu+1},\\
&&T_{||}-T_{||0}\propto \sin\theta[P(1+P)-(1-P)(1+P\cos\theta)] V^{\nu+2},\label{eq:19}\\
&&T_{\bot}-T_{\bot0}\propto P\phi\sin\theta(1-\cos\theta) |V|^{\nu+2}.\label{eq:20}
\end{eqnarray}
The detail of these expressions can be found in Ref. \onlinecite{magnon}. Interestingly the perpendicular torque and the conductance both acquire a component that is symmetric against the bias. Furthermore, since the electron-magnon interaction mixes the majority and minority channels, the angular dependence is also affected, contrary to the case of electron-phonon coupling.\par

The finite temperature situation has been studied in Ref. \onlinecite{magnon} and gives rise to a non-linear dependence as a function of both voltage and temperature. Actually, competing mechanisms take place when both magnon emission and absorption are accounted for. Let us consider the torque exerted on the right electrode magnetization. Magnon emission (absorption) occuring at the left interface increases (reduces) the effective spin-polarization of the incoming electrons, therefore enhancing (lowering) the spin torque exerted on the right electrode. Symmetrically, electron-magnon interactions occuring at the right interface also affects the effective polarization of electrons coming from the right reservoir. Finally, we must stress out that the detailed temperature and bias dependencies presented here are strongly conditioned by the electrons, phonons and magnon band structures.\par

\subsection{Bulk scattering}

In contrast with interfacial scattering, in the case of
bulk scattering (by impurities or magnons) the spin torque is no more described by the purely
interfacial spin current since spin relaxation can not be neglected
in the bulk of the layers. Therefore, the spin torque reads:
\begin{equation}
{\bf T}=\int_\Omega  \frac{J}{\hbar}{\bf m}\times{\bf
M}d\Omega=\int_\Omega [-\nabla{\cal J}_s-\frac{{\bf m}}{\tau_{sf}}]d\Omega
\end{equation}
The presence of a finite spin relaxation time ($1/\tau_{sf}\neq0$)
induces a coupling between the two components of the spin torque, so
that the perpendicular torque now involves a contribution of both
in-plane and perpendicular interfacial spin current densities. In a MTJ, the interfacial densities of state are usually only affected by the first few monolayers away from the interface. Therefore, since the spin-diffusion length is on the order of 5-15nm, we can assume that the tunneling process is almost not affected by the presence of spin-flip scattering. Then, the interfacial spin current can be identified to the spin torque without spin-flip: ${\cal J}_s(x=0)={\bf T}_0$. As a consequence, the actual spin torque has the form:

\begin{eqnarray}
T_{||}&=&T_{||0}+\frac{\tau_J}{\tau_{sf}}T_{\bot0}\\
T_{\bot}&=&T_{\bot0}-\frac{\tau_J}{\tau_{sf}}T_{||0}
\end{eqnarray}
where $\tau_J=\hbar/J$. In the case of a symmetric magnetic tunnel junction in the absence of interfacial inelastic scattering, $T_{||0}$ and $T_{\bot0}$ are given by Eqs. (\ref{eq:SEip})-(\ref{eq:SEop}). For low bias voltage, when the spin-flip is dominated by Elliott-Yafet spin scattering, $\tau_{sf}$ is bias-independent (but temperature dependent) and the perpendicular torque gains a linear component $a_1V\tau_J/\tau_{sf}$.\par

At large bias, or non-zero temperature, the spin-flip scattering is dominated by the electron-magnon interaction. As showed by Li et al. \cite{deac}, the spin-flip relaxation time due to electron-magnon interaction is inversely proportional to $|V|$. Consequently, the presence of bulk magnons results in an additional component in the perpendicular torque of the form $\propto j_e|V|$. This is in sharp contrast with the case of interfacial magnons, where the additional component is simply $|V|$.

\section{Conclusion\label{s:con}}

In summary, we studied the influence of structural asymmetries and inelastic tunneling on the bias dependence of the spin transfer torque in MTJs, using either the free electron model or the Transfer Hamiltonian formalism. Our results are summarized below:
\begin{enumerate}
	\item Structural asymmetries: the perpendicular torque and the conductance acquire a {\em antisymmetric} linear component of the form $\propto V$, while the bias dependence of the in-plane torque is still described by Eq. (\ref{eq:SEip}).
	The obtained formulae provide consistent results in the low-bias
region and at low band filling with the numerical results of the
tight binding model \cite{kioussis} and are in good agreement with the numerical
results of the free electron model \cite{xiao,wil}. Consequently, they can serve as a guideline to design the spin torque bias dependence, as demonstrated by Oh et al. \cite{oh}.\par
	\item Inelastic interfacial scattering: the symmetry of the MTJ is conserved and the perpendicular torque and conductance acquire a {\em symmetric} linear component of the form $\propto |V|^n$, $n\in\mathbb{N}$. The presence of magnons or phonons interactions is usually revealed through peaks in the conductance derivative. The influence of the temperature has been briefly discussed.
	\item Bulk spin-flip scattering: the spin torque is no more equal to the net transfer of angular momentum. The relaxation of the spin accumulation induces a mixing between the two components of the torque, giving rise to an {\em antisymmetric} component of the form $V$ and $j_e|V|$ in the case of impurity- and magnon-induced spin scattering, respectively. Since the resistance is dominated by the barrier, the contribution of bulk scattering is usually negligible on the conductance.
\end{enumerate}

Finally, we suggest that a link exists between the signature of asymmetry and inelastic scattering in the perpendicular torque and conductance. Since both are symmetric against the bias in a symmetric MTJ, the introduction of structural asymmetries or inelasticity affects both quantities, but in different ways. The careful analysis of the perpendicular torque together with the conductance should give important clues on the origin of the additional linear terms, as suggested in Ref. \onlinecite{oh} in the case of structural asymmetries. Note however, that the conductance remains unaffected by bulk scattering and therefore, the influence of bulk magnons cannot be analyzed by comparing the perpendicular torque and the conductance.

\begin{acknowledgments}
This work was supported by NSF (DMR-0704182)
and DOE (DE-FG02-06ER46307). A. M. acknowledges fruitful discussions with M. Chshiev.
\end{acknowledgments}

\appendix
\section{Wave functions in the large barrier approximation}
We use the free electron model within the Keldysh formalism as described in Ref. \onlinecite{manchon}. The electron wave vectors for majority and minority spin in the left and right electrodes and in the barrier are then:
\begin{eqnarray}
&&k_{1,2}=\sqrt{\frac{2m}{\hbar^2}\left(E-E_{||}\pm J_L-\frac{eV}{2}\right)}\\
&&k_{3,4}=\sqrt{\frac{2m}{\hbar^2}\left(E-E_{||}\pm J_R+\frac{eV}{2}\right)}\\
&&\kappa=\sqrt{\frac{2m}{\hbar^2}\left(\phi_L+\frac{eV}{2}+E_F-E+E_{||}-\frac{x}{d}(eV-\Delta\phi)\right)}
\end{eqnarray}
The indices 1,3 (2,4) refer to the majority (minority) spin. The wave function of an electron of injected from the i-th electrode with an initial spin $\sigma$ is represented in the vector form $\Psi_{\sigma}^i=(\Psi_{\uparrow\sigma}^i,\Psi_{\downarrow\sigma}^i)$. The wave functions for the electrons from the left and right electrode at the interfaces are then \cite{manchon}:
\begin{eqnarray}
&&\Psi_{\uparrow\uparrow}^L=\frac{\sqrt{2k_1}}{k_1+i\kappa_1}\\
&&\Psi_{\downarrow\uparrow}^L=4\sqrt{2k_1}\frac{\kappa_1\kappa_2(k_3-k_4)}{den}\sin\theta\\
&&\Psi_{\downarrow\downarrow}^L=\frac{\sqrt{2k_2}}{k_2+i\kappa_1}\\
&&\Psi_{\uparrow\downarrow}^L=4\sqrt{2k_2}\frac{\kappa_1\kappa_2(k_3-k_4)}{den}\sin\theta\\
&&\Psi_{\uparrow\uparrow}^R=4iE_n\frac{\sqrt{2k_3\kappa_1\kappa_2}}{den}(k_2+i\kappa_1)(k_4+i\kappa_2)\cos\frac{\theta}{2}\\
&&\Psi_{\downarrow\uparrow}^R=4iE_n\frac{\sqrt{2k_3\kappa_1\kappa_2}}{den}(k_1+i\kappa_1)(k_4+i\kappa_2)\sin\frac{\theta}{2}\\
&&\Psi_{\downarrow\downarrow}^R=4iE_n\frac{\sqrt{2k_4\kappa_1\kappa_2}}{den}(k_1+i\kappa_1)(k_3+i\kappa_2)\cos\frac{\theta}{2}\\
&&\Psi_{\uparrow\downarrow}^R=-4iE_n\frac{\sqrt{2k_4\kappa_1\kappa_2}}{den}(k_2+i\kappa_1)(k_3+i\kappa_2)\sin\frac{\theta}{2}
\end{eqnarray}
with $den=2E_n^2(k_1+i\kappa_1)(k_2+i\kappa_1)(k_3+i\kappa_2)(k_4+i\kappa_2)$, $E_n=\exp\left[-d\sqrt{\frac{2m}{\hbar^2}}\int_0^d\kappa dx\right]$ is the exponential factor and $\kappa(x=0)=\kappa_1$,$\kappa(x=d)=\kappa_2$. The above equations together with the integration rules mentioned in section II.A. are sufficient to describe the transport properties of the junction.

\section{Currents and Torques in the large barrier approximation}
By definition, the charge and spin currents are defined as 
\begin{eqnarray}
J_i=\frac{2e}{h}\int\int dE d{\bf k}_{||}\left(\langle\sigma_i\otimes\nabla\rangle_L f_L+\langle\sigma_i\otimes\nabla\rangle_R f_R\right),
\end{eqnarray}
where $i=0,x,y,z$ and ${\bm \sigma}=(\sigma_x,\sigma_y,\sigma_z)$ are the spin Paul matrices, $\sigma_0$ is the identity and $\langle...\rangle_{L,R}$ denotes quantum mechanical averaging, involving the rightward and leftward spin-dependent wave functions defined in Ref. \onlinecite{manchon}. $f_L$ and $f_R$ are the Fermi distribution functions of the left and right reservoirs. Expanding these wave functions up to the lowest order in the barrier height, the charge and spin currents for a majority electron issued from the left reservoir are:
\begin{widetext}
\begin{eqnarray}
&&J_{eL}^{\uparrow}=\frac{2e}{h}\int\int dE d{\bf k}_{||}\frac{8k_1\kappa_1\kappa_2(k_3+k_4)(\kappa_1^2+k_2^2)(\kappa_2^2+k_3k_4)}{(\kappa_1^2+k_1^2)(\kappa_1^2+k_2^2)(\kappa_2^2+k_3^2)(\kappa_2^2+k_4^2)}[1+P_L\cos\theta]f_L\\
&&T_{||L}^{\uparrow}=\int\int dE d{\bf k}_{||}\frac{4k_1\kappa_1\kappa_2(k_3-k_4)(\kappa_1^2+k_2^2)(\kappa_2^2-k_3k_4)}{(\kappa_1^2+k_1^2)(\kappa_1^2+k_2^2)(\kappa_2^2+k_3^2)(\kappa_2^2+k_4^2)}f_L\sin\theta\\
&&T_{\bot L}^{\uparrow}=\int\int dE d{\bf k}_{||}\frac{4k_1\kappa_1\kappa_2^2(k_3^2-k_4^2)(\kappa_1^2+k_2^2)}{(\kappa_1^2+k_1^2)(\kappa_1^2+k_2^2)(\kappa_2^2+k_3^2)(\kappa_2^2+k_4^2)}f_L\sin\theta
\end{eqnarray}
\end{widetext}
and $P_L=\frac{(k_1-k_2)(\kappa_1^2-k_1k_2)}{(k_1+k_2)(\kappa_1^2+k_1k_2)}$ is Slonczewski's polarization \cite{slonc89}.The contribution for a minority electron is obtained by performing the following replacements: $k_{1,3}\leftrightarrow k_{2,4}$ and $\theta\rightarrow-\theta$. Similarly, the contribution of electrons issued from the right reservoir is obtained by performing the following replacements: $\kappa_1\leftrightarrow\kappa_2$, $(1,2)\leftrightarrow(3,4)$ and $f_L\rightarrow f_R$. The final expressions are then:
\begin{eqnarray}
&&J_{e}=J_{eL}^{\uparrow}+J_{eL}^{\downarrow}-J_{eR}^{\uparrow}-J_{eR}^{\downarrow}\\
&&T_{||}=T_{||L}^{\uparrow}+T_{||L}^{\downarrow}-T_{||R}^{\uparrow}-T_{||R}^{\downarrow}\\
&&T_{\bot}=T_{\bot L}^{\uparrow}-T_{\bot L}^{\downarrow}+T_{\bot R}^{\uparrow}-T_{\bot R}^{\downarrow}
\end{eqnarray}
\section{Analytical expressions for Spin Torques and Conductance}
After some algebra using Eqs. (\ref{eq:analtip})-(\ref{eq:analg}), we obtain the following results:
\begin{widetext}
\begin{eqnarray}\label{eq:22}
T_{||0}&=&\frac{\hbar^2}{2m}\frac{\kappa_0^6}{2\pi^2\beta}\frac{(k_\uparrow^2-k_\downarrow^2)(\kappa_0^4-k_\uparrow^2k_\downarrow^2)}{(\kappa_0^2+k_\uparrow^2)^2(\kappa_0^2+k_\downarrow^2)^2}e^{-2\beta}\\\label{eq:23}
T_{\bot0}&=&\frac{\hbar^2}{2m}\frac{\kappa_0^7}{\pi^2\beta^2}\frac{(k_\uparrow^2-k_\downarrow^2)(k_\uparrow-k_\downarrow)(\kappa_0^2-k_\uparrow^2k_\downarrow^2)}{(\kappa_0^2+k_\uparrow^2)^2(\kappa_0^2+k_\downarrow^2)^2}e^{-2\beta}\\\label{eq:24}
G^{p,ap}_0&=&\frac{2e^2}{\hbar}\frac{\kappa_0^4}{\pi^2\beta}\frac{(k_\downarrow+k_\uparrow)^2(k_\uparrow k_\downarrow+\kappa_0^2)^2\pm(k_\downarrow-k_\uparrow)^2(k_\uparrow k_\downarrow-\kappa_0^2)^2}{(\kappa_0^2+k_\uparrow^2)^2(\kappa_0^2+k_\downarrow^2)^2}e^{-2\beta}
\end{eqnarray}
\end{widetext}
and
\begin{widetext}
\begin{eqnarray}
a_1&=&1+\frac{\kappa_0^2[2(\beta-1)k_\uparrow^2k_\downarrow^2+\kappa_0^2(k_\uparrow^2+k_\downarrow^2)]}{2k_\uparrow k_\downarrow ((\beta-1)k_\uparrow^2k_\downarrow^2-(\beta-3)\kappa_0^4+(k_\uparrow^2+k_\downarrow^2)\kappa_0^2)}\frac{\Delta\phi}{\phi}\nonumber\\
&&-\frac{2(\beta-1)k_\uparrow^2k_\downarrow^2(k_\uparrow^2+k_\downarrow^2)+5\kappa_0^2(k_\uparrow^4+k_\downarrow^4)-2k_\uparrow^2k_\downarrow^2\kappa_0^2}{8k_\uparrow k_\downarrow ((\beta-1)k_\uparrow^2k_\downarrow^2-(\beta-3)\kappa_0^4+(k_\uparrow^2+k_\downarrow^2)\kappa_0^2)}\frac{\Delta J}{\phi}\\
a_2&=&\frac{\kappa_0^2[2(\beta-1)k_\uparrow^2k_\downarrow^2-\kappa_0^2(k_\uparrow^2+k_\downarrow^2)](\kappa_0^2+k_\uparrow^2)(\kappa_0^2+k_\downarrow^2)}{4k_\uparrow^3 k_\downarrow^3 ((\beta-1)k_\uparrow^2k_\downarrow^2-(\beta-3)\kappa_0^4+(k_\uparrow^2+k_\downarrow^2)\kappa_0^2)}-\frac{\beta}{24}\frac{\Delta\phi}{\phi}\nonumber\\
&&-\frac{k_\uparrow^4(\kappa_0^2+k_\uparrow^2)^2((\beta-1)k_\downarrow^2-\kappa_0^2)+k_\downarrow^4(\kappa_0^2+k_\downarrow^2)^2((\beta-1)k_\uparrow^2-\kappa_0^2)}{8k_\uparrow^4 k_\downarrow^4 ((\beta-1)k_\uparrow^2k_\downarrow^2-(\beta-3)\kappa_0^4+(k_\uparrow^2+k_\downarrow^2)\kappa_0^2)}\frac{\Delta J}{J}\\
b_1&=&\frac{\beta-2}{8}\frac{k_\uparrow k_\downarrow(k_\uparrow k_\downarrow-3\kappa_0^2)(k_\uparrow-k_\downarrow)-(2k_\uparrow k_\downarrow-\kappa_0^2)(k_\uparrow^3-k_\downarrow^3)+(k_\uparrow^5-k_\downarrow^5)}{k_\uparrow k_\downarrow(k_\uparrow-k_\downarrow)(\kappa_0^2-k_\uparrow k_\downarrow)}\frac{\Delta J}{\phi}\nonumber\\
&&-\frac{k_\uparrow k_\downarrow(\beta+3)+(2\beta-9)\kappa_0^2}{6(k_\uparrow k_\downarrow-\kappa_0^2)}\frac{\Delta\phi}{\phi}\\
b_2&=&\beta\left(\frac{\beta}{8}-\frac{5}{12}\right)\\
g^p_1&=&\frac{\kappa_0^4}{2k_\uparrow^2 k_\downarrow^2}\frac{k_\uparrow^2(\kappa_0^2+k_\uparrow^2)^2-k_\downarrow^2(\kappa_0^2+k_\downarrow^2)^2}{k_\uparrow^2(\kappa_0^2+k_\downarrow^2)^2+k_\downarrow^2(\kappa_0^2+k_\uparrow^2)^2}\frac{\Delta J}{\phi}-\left(\frac{\beta}{12}-\frac{3}{8}\right)\frac{\Delta \phi}{\phi}\\
g^{ap}_1&=&\frac{\kappa_0^2(k_\uparrow^2- k_\downarrow^2)^2[k_\uparrow^2(3k_\downarrow^2+\kappa_0^2)(k_\uparrow^2+\kappa_0^2)+k_\downarrow^2(3k_\uparrow^2+\kappa_0^2)(k_\downarrow^2+\kappa_0^2)]}{16k_\uparrow^4k_\downarrow^4(\kappa_0^2+k_\uparrow^2)(\kappa_0^2+k_\downarrow^2)}\frac{\Delta J}{J}-\frac{\beta}{12}\frac{\Delta\phi}{\phi}\\
g^{p,ap}_2&=&\frac{\beta}{8}(\beta-1)
\end{eqnarray}
\end{widetext}
where $k_{\uparrow,\downarrow}=\sqrt{\frac{2m}{\hbar^2}(E_F\pm J)}$ and $\kappa_0=\sqrt{\frac{2m}{\hbar^2}\phi}$. When the barrier becomes thinner, a corrective multiplication factor
of the form $(1+\frac{3}{2\beta}+\frac{3}{4\beta^2})$ should be
inserted into Eqs. (\ref{eq:22})-(\ref{eq:24}). Note that $T_{\bot0}$ and
$G_0$ are similar to previous derivations using a free electron
model \cite{slonc89,brinkman}. The above relations are limited to low bias voltage in low band filling systems. Using more realistic densities of states, these relations may by modified.

\end{document}